\documentclass[prl,twocolumn,showpacs,showkeys]{revtex4}

\usepackage{amssymb, amsmath, amsfonts, latexsym, verbatim, mathrsfs}

\usepackage{graphicx}

\begin{document}

\title{About possible extensions of quantum theory}
\thanks{We are grateful to S. Goldstein and D. Duerr for a critical reading of the paper. R. R. acknowledges support by the ARO MURI grant W911NF-11-1-0268.}

\author{GianCarlo Ghirardi}
\email{ghirardi@ictp.it}
\affiliation{Department of Physics, University of Trieste, and the Abdus Salam ICTP, Trieste (Italy)}
\author{Raffaele Romano}
\email{rromano@iastate.edu}
\affiliation{Department of Mathematics, Iowa State University, Ames, IA (USA) }

\begin {abstract}
\noindent Recently it has been claimed that {\it no extension of quantum theory can have improved predictive power}, the statement following, according to the authors, from the assumptions of free will and of the correctness of quantum predictions concerning the correlations of measurement outcomes. Here we prove that the argument is basically flawed by an inappropriate use of the assumption of free will. In particular, among other implications, the claim, if correct, would imply that Bohmian Mechanics is incompatible with free will. This statement, appearing in the paper, derives from  the unjustified identification of free will with the no-signaling constraint and of a
 purely formal and not physical use of such a constraint.
\end{abstract}


\pacs{03.65.Ta, 03.65.Ud}

\keywords{Signaling, Free will, Quantum Theory}

\maketitle


\section{1. INTRODUCTION}
In a recent paper, Colbeck and Renner~\cite{colbeck} have argued that  the  assumptions
\begin{itemize}
\item that the observers can freely choose  which specific measurements to perform on the constituents of a composite quantum system in an entangled state, and
\item that the statistics of the  outcomes agrees with quantum predictions,
\end{itemize}
imply that any extension of quantum theory cannot convey any additional information about the outcomes of future measurements  with respect to the one supplied by  the state vector. In other words, no completion of the theory is possible.

The authors consider two particles heading towards two measurement devices that have a number of possible settings, identified by two parameters A and B,  that can be chosen freely and independently by the experimenters. The measurements yield results denoted by X and Y, respectively. At this point the authors raise the question of whether there could exist a further source of information $\Xi$ allowing to  predict more accurately  the outcomes.

The novelty of the approach with respect, e.g., a standard hidden variable completion of quantum mechanics, stays in  the fact that the authors do not require (but they even do not exclude)  that such information yields a precise knowledge of the outcomes (determinism) and in the fact that they do not assume that the further information be encoded in a classical system (as it happens in the case of standard hidden variable theories).  In particular, it might happen that to get the additional information one has to access the source by a typical quantum procedure, i.e. to input a quantum system C (a sort of measuring apparatus) and to use the output Z from C to increasee his knowledge concerning the state of the system A+B. The only crucial aspect addressed in Ref.~\cite{colbeck} is whether the further information $\Xi$ might enable one to make more accurate predictions concerning the outcomes than those implied by the knowledge of the state vector.
Moreover, the authors specify that A,B,C and X,Y,Z are spacetime random variables whose values can be associated to precise spacetime points $(t, r_{1}, r_{2}, r_{3})$. This important specification is  useful to make fully precise the assumption that the two measurement procedures and the associated outcomes are space-like with respect to one another.

At this point the authors formulate their first request, denoted as FR, that the observers can independently choose the measurements they perform. Using their words:
\begin{quote}
{\it Assumption FR is that the input, A, of a measurement process can be chosen such that it is uncorrelated with certain other spacetime random variables, namely all those whose coordinates lie outside the future lightcone of the coordinates of A.}
\end{quote}
It is meant that this assumption holds also for B and C.
The second fundamental request (denoted as QM) is that the present quantum theory is correct, i.e.:

\begin{quote}
{\it Measurement outcomes obey quantum statistics and [...]  all processes within quantum theory can be considered as unitary evolutions, if one takes into account the environment. [Moreover] the second part of the assumption need only hold for microscopic processes on short timescales.}
\end{quote}

The main mathematical objects  the paper deals with are joint distributions of measurements and outcomes, typically objects like $P_{XYZ|ABC}$, $P_{YZ|ABC}$, $P_{XY|AB}$ and similar ones.
The authors start by formalizing their request FR of free choice of the settings A, B and C in the following precise way:
\begin{eqnarray}\label{FR}
P_{A|BCYZ} &=& P_{A} \nonumber \\
P_{B|ACXZ} &=& P_{B} \\
P_{C|ABXY} &=& P_{C}. \nonumber
\end{eqnarray}

From such relations they derive, with simple manipulations involving conditional and unconditional probabilities, the three following relations:
\begin{eqnarray}\label{NS}
P_{YZ|ABC}&=&P_{YZ|BC} \nonumber \\
P_{XZ|ABC}&=&P_{XZ|AC} \\
P_{XY|ABC}&=&P_{XY|AB}, \nonumber
\end{eqnarray}
which are the non-signaling constraints on which all the subsequent arguments are based.
In fact these constraints, when applied to appropriately chosen measurements on a maximally entangled state (or, by extending the procedure, to an arbitrary entangled state) lead to the basic conclusion of the paper which the authors summarize in the following way:
the extra information $\Xi$ is of no use since {\it the distribution of X given A and $\Xi$ is the same as the distribution of X given only A}. Accordingly, the authors claim to have proved that {\it under the assumption that measurement settings can be chosen freely, quantum theory really is complete}.

In our opinion, this conclusion is incorrect because the formal expression (1) of the FR assumption  in~\cite{colbeck} does not properly express the free will condition, and additional constraints are hidden in it. Our wariness comes first of all from the consideration that hidden
variables models predictively equivalent to quantum mechanics, where additional knowledge on the state of the system can modify the quantum statistics without allowing faster than light signalling and in which free will is tacitly assumed, have been developed. In fact, in these theories the main issue is the {\it accessibility} of the hidden variables, and free will is not called into question. The prominent example of such theories is Bohmian mechanics, where the hidden variables are given by the positions of the particles, which are assumed to be fully unaccessible, and the experimenters have free will. For sake of clearness and for future reference we will make to it, in Appendix A we describe an elementary hidden variable model  for the singlet state of a pair of two-levels systems, which, for the present purposes, is conceptually equivalent to Bohmian mechanics and which has been recently discussed by us \cite{ghirardi2} with reference to a proposal by Leggett \cite{leggett}. It is consistent with  quantum theory, it violates the conditions (1)  despite the fact that observers can very well be assumed to have free will, and, in it, the accessible additional information on the state leads to probabilities which differ from those of quantum mechanics.

The result of Colbeck and Renner is valid for both deterministic or probabilistic completions of quantum mechanics. Nonetheless, in order to contextualize this contribution and to clearly formulate our criticism to the authors' assumptions, in Section 2 we mainly concentrate on deterministic models, like Bohmian mechanics and the modified Bell model presented in the Appendix. We show that, for such a class of models, the result derived in~\cite{colbeck} is not new, although usually it is interpreted in a different way. In Section 3 we address the problem of the accessibility of the hidden variables and its relations with the non-signalling conditions. In Section 4 we discuss the standard approach to free will in deterministic theories, and comment upon the assumption FR under this perspective.
In Section 5 we propose a possible alternative  approach to the problem of imposing  free will which makes explicitly clear that the constraint (1) is stronger than necessary. In particular, under this perspective, we show that it is possible to have theories in which the free will assumption is respected, but the non-signalling conditions are violated. In Section 6 we outline the analogies between the theorem presented in~\cite{colbeck} and  the famous example of the von Neumann no-go theorem against deterministic hidden variable theories. Finally, we summarize our results and conclude in Section 7.



\section{2. ON DETERMINISTIC COMPLETIONS OF QUANTUM MECHANICS}
We are  confronted with the following situation: we have an extremely simple model \cite{ghirardi2} (a trivial generalization of a proposal \cite{bell} by Bell) and the very general theory worked out by Bohm, which  everybody has taken seriously as perfectly acceptable completions of quantum mechanics consistent with the free will assumption, and  both of them violate  the constraints that are put at the very basis of the argument of Ref.~\cite{colbeck}. Since such constraints derive in a straightforward way from (\ref{FR}), which the authors consider as appropriately expressing the physical request that the settings of the apparatuses can be chosen freely, it is just this condition which has to be critically investigated.

As pointed out by Colbeck and Renner themselves, the non-signaling constraints (\ref{NS}) imply $P_{X|ABZ} = P_{X|AZ}$ and $P_{Y|ABZ} = P_{Y|BZ}$, i.e., they require the candidate theories under consideration  to satisfy the Parameter Independence (PI) condition. Now, it is universally known  that any deterministic theory which reproduces the predictions of quantum mechanics cannot violate Outcome Independence (OI) and therefore, since it must violate Bell's locality, it must violate precisely the PI request. From this point of view the whole and complicated proof of Ref.~\cite{colbeck} becomes trivial in the case of deterministic completions of quantum mechanics. By indicating as $\{Det. \; Compl.\}$ any deterministic  completion of the theory, the whole argument, for the case under discussion, can be summarized in the following  terms:
\begin{eqnarray}
&&FR\supset PI \Longleftrightarrow \neg  (PI)\supset \neg FR \\
&&\{Det. \; Compl.\}\supset \neg (PI)\supset \neg FR \nonumber,
\end{eqnarray}

\noindent i.e., as already stated,
\begin{equation}
FR\supset \neg \{Det. \; Compl.\}.
\end{equation}

Actually, the just outlined derivation of the conclusion (4) for deterministic hidden variable theories turns out to have a much more general validity. In fact, A. Fine \cite{fine} has been able to prove that for any factorizable stochastic hidden-variable model for a correlation experiment there exist a deterministic hidden variable model for the same experiment. Thus, the straightforward derivation of the impossibility of a deterministic completion following from (1) actually applies also to a much larger, and actually  the most physically interesting and investigated, class of stochastic hidden variable models. We believe that it is appropriate to quote Fine himself on this point:

\begin{quote}
{\it despite appearances, no significant generality is achieved in moving from deterministic hidden variables to stochastic ones ... [and] ... no significant generality is achieved by those derivations of the Bell/CH inequality that dispense with explicit reference to hidden variables and/or determinism.}

\end{quote}

In full agreement with their identification of the FR with the relations (1), Colbeck and Renner explicitly state that {\it in the context of de Broglie-Bohm theory the presence of nonlocal hidden variables contradicts assumption FR}, and, as a consequence, in this theory there is no free will.

This quite general and revolutionary statement seems to us fully inappropriate and it gives clear hints concerning the reasons for which the analysis of Ref.~\cite{colbeck} is not conclusive. In our opinion the weak points of the argument are: (i) the assumption that free will is strictly connected to the causal structure of the problem, and, in particular, the request that free will implies impossibility of superluminal communication; (ii) a non clear position concerning the accessibility or inaccessibility of the further information completing the one given by the state vector. It seems to us that it has been just the focus the authors have put on quantum non locality and the associated problem of the potential ``spooky action at-a-distance" that has led them \footnote{Actually, the authors of Ref.~\cite{colbeck} have felt the necessity, in the Supplementary Information to their paper, to make precise remarks on the notion of nonlocality and its connections with the assumption of free will.} to express the free will request in the form (1).

In the next section we will comment, first of all, on the problem of the relations between faster than light signaling and the accessibility of the hidden variables.

\section{3. FREE WILL AND THE ACCESSIBILITY OF THE HIDDEN VARIABLES}

Colbeck and Renner, in their paper, have not faced the crucial problem of the accessibility or inaccessibility of the further information on the system, for brevity sake, let us say of the hidden variables~\footnote{The only point of the paper from which one can get a hint concerning the accessibility of the hidden variables is the assumption that the authors make concerning the static nature of the additional information. This choice implies that one can consider of getting it at any time, in particular at a time which makes $Z$ and $C$ space-like with respect to $A$, $X$ and $B$, $Y$. We note that this fact is intended to put $C$ and $Z$ on the same footing as the other variables, so that, e.g., (1) and (2) deal with $C$ precisely in the same manner they deal with $A$ and $B$. But this raises a non trivial problem: if $C$ and $Z$ are space like with respect to the choices and the detections in $A$ and $B$, they turn out to be, by assumption, inaccessible to the experimenters at the two wings of the apparatus. But then, since the problem of faster than light signaling concerns precisely these experimenters, what is the use for them of the existence of the supplementary information $\Xi$? }. Actually,  there is no doubt that a theory completing quantum mechanics by controllable hidden variables which (by themselves or used in addition to the state vector) determine the outcomes of all conceivable observables, allows faster than light communication between observers. But when, as in the case of Bohmian mechanics, the hidden variables are assumed to be fundamentally unaccessible, in the sense that  the two experimenters have no control on them, superluminal communication becomes impossible and there is no need and no physical reason to impose, at the basic level, the non-signaling conditions in the precise form (2). With reference to this point one can see the detailed analysis by A. Shimony~\cite{shimony2} concerning controllable and uncontrollable nonlocality. Since the requests (1) imply the NS relations (2), a violation of such relations implies a rejection of condition (1).

We are firmly convinced that this is the crucial point of the problem: the compatibility of any completion of quantum mechanics with the relativistic requirement of no superluminal communication has much more to do with the accessibility or nonaccessibility of the formal elements completing the theory than with the possibility of freely choosing the apparatus settings. It is in this spirit that we do not agree with Colbeck and Renner's way of imposing free will.

For further reference, we denote by $\Lambda$ the so-called {\it ontic state} of the system, that is, the most complete specification of the state of the system which is {\it in principle} possible within a given theory. In general, it embodies also the information contained in the quantum state vector, and it differs from the hidden variables (denoted by $\lambda$ in this work), which are usually assumed to complete the information given by the state vector. We stress that $\Lambda$ is generally assumed (as it is necessary, see below) to be not fully accessible, and this fact allows to circumvent the problem of faster than light communication. Now, it is not clear whether, in Ref.~\cite{colbeck}, the additional information $\Xi$ jointly with the quantum state vector is equivalent to knowledge of $\Lambda$, or it is rather expected to represent only the accessible information contained in $\Lambda$. Nonetheless, the operational procedure for accessing  $\Xi$ suggests that the authors  assume that the accessible information is given by $Z$.
\section{4. FREE WILL AND DETERMINISM}

In contrast with the authors' position, in the usual EPR-Bohm type of scenario one usually (and tacitly) assumes that the settings of the devices can be chosen ``freely", i.e., independently of the specification of the state of the incident particles. Typically, in the most significative experiments (as those of Aspect) the settings are generated by independent random number generators, so that there is nothing in the past light cone of the individual measurement events  determining which choice will be made.

However, in fully deterministic theories such as Bohmian mechanics itself the choice of the settings might be (in principle) influenced by events in their past and, therefore, the pasts of the settings overlap with the situations which will play a role concerning the states of the particles being measured prior to the measurement processes. A complete specification of the state will therefore include facts which in some measure are relevant for the settings, so that there is a certain subtle incompatibility of the assumption that the choice of the settings can be made independently of the state of the particles. It seems that it is just this point which has led Colbeck and Renner to be so strict in putting forward their request FR. However, the potential tension between possible past influences on the settings and the assignment of the state of the physical system has already been exhaustively discussed and clarified in the literature.

In particular, this problem has been   considered both by Bell and by Shimony, Clauser and Horne~\cite{shimony} and has been recently lucidly analyzed by T. Norsen~\cite{norsen} along lines which we recall here. To overcome the just mentioned difficulty one merely allows that the events in the relevant region can be divided into disjoint classes: those which are influenced by the preparation procedure and specify the state  of the system and those which enter in the choice of the settings. These two classes are far from jointly exhaustive so that one can expect a causal distance between them which makes the ``freedom" assumption a quite reasonable one. As Bell himself has acknowledged~\cite{bell2}:
\begin{quote}
{\it it is not permissible to regard the experimental settings in the analyzers as independent of the supplementary variables $\lambda$, in that $a$ and $b$ could be changed without changing the probability distribution $\rho(\lambda)$.
Now, even if we have arranged that $a$ and $b$ are generated by apparently random radioactive devices, housed in separate boxes and thickly shielded, or by Swiss national lottery machines, or by elaborated computing programmes, or by apparently free will experimental physicists, or by some combination of all of these, we cannot be {\it sure} that [the settings] $a$ and $b$ are not significantly influenced by the same factors $\lambda$ that influence [the outcomes] A and B. }
\end{quote}
\noindent but he has also felt the necessity of  making absolutely clear that:
\begin{quote}
{\it ... this way of arranging quantum mechanical correlations would be even more mind boggling that one in which causal chains go faster than light. Apparently, separate parts of the world would be deeply and conspiratorially entangled, and our apparent free will would be entangled with them.}
\end{quote}
Moreover he subsequently  added~\cite{bell3}:
\begin{quote}
{\it One can envisage then theories in which there  just {\it are} no free variables for the polarizer angles to be coupled to. In such `super deterministic' theories the apparent free will of the experimenter, and any other apparent randomness, would be illusory. Perhaps such a theory could be both locally causal and in agreement with quantum mechanical predictions. However I do not expect to see a serious theory of this kind. I would expect a serious theory to permit `deterministic chaos' or `pseudo randomness', for complicated subsystems (e.g. computers) which would provide variables sufficiently free for the purpose at hand. But I do not have a theorem about that.}
\end{quote}
It is also useful to recall, as Norsen did, the clear-cut position taken in Ref.~\cite{shimony} on the issue of the ``freedom" or ``no conspiracies" assumption we have just mentioned:
\begin{quote}
{\it we feel that it is wrong on methodological grounds to worry seriously about [such an assumption] if no specific causal linkage is proposed. In any scientific experiment in which two or more variables are supposed to be randomly selected, one can always conjecture that some factor in the overlap of the backward light cones has controlled the presumably random choices. But, we maintain, skepticism of this sort will essentially dismiss all results of scientific experimentation. Unless we proceed under the assumption that hidden conspiracies of this sort do not occur, we have abandoned in advance the whole enterprise of discovering the laws of nature by experimentation.}
\end{quote}
This kind of argument is particularly pertinent and significative when one is considering theories which violate PI, but  in which the hidden variables are assumed not to be fully accessible.

Notice that, under this perspective, the request embodied in the assumption FR appears too strong even only from a conceptual point of view, irrespective of its mathematical formulation. What really matters is that the variables in the intersection of the past light cones of the events corresponding to the measurement choices, cannot correlate the choices of the two far away experimenters. But separate influences could be accepted, while they are prevented by FR: expressing free will through FR can be questioned also on this basis.

The mistake of Colbeck and Renner should now be clear: even if the unavoidable nonlocal character of the theories we are dealing with is taken into account, it is not physically reasonable to impose the free will condition in the way chosen by the authors. Actually, there is no compelling physical reason to impose the constraints (1), which in general cannot be identified with the free will requirement, so that extensions of quantum theory with improved predictive power can exist, at least in principle (for instance, see~\cite{ghirardi2}).

Before concluding this Section, we consider it appropriate to enrich the above analysis based on general logical and epistemological considerations with an example strictly related to the model we have presented in the appendix. As we have formulated it, the model is perfectly neutral concerning the problem of the choice of the settings. However, one could pretend that, given the framework we have chosen - specifically one in which the completion of the theory implies determinism - we should also include the apparatuses and their settings in the game, so that also the choices of the settings would be predetermined within the enlarged theoretical scheme.

Apart from the not fully compelling direct extension of the proposal from microsystems to macroscopic apparatuses, let us take the challenge represented by this remark. Our system is now composed of 3 subsystems, the apparatuses A and B and the composite system in the state $\psi$. In the spirit of our approach we might then consider the initial apparatus states $\Psi^{(A)}$ and $\Psi^{(B)}$, further variables $\lambda_{A}$ and $\lambda_{B}$ supplementing the information given by the corresponding state vectors and, finally, the state $\psi$ and the variable $\lambda$ of the scheme. We note that this separate consideration of the various state vectors and hidden variables is not  arbitrary, actually it is quite natural, in particular when  the initial state  is non entangled. In such a case, in quantum mechanics the total state vector would factorize, $\Psi_{total} = \Psi_{A} \otimes \Psi_{B} \otimes \psi$. Accordingly, the probabilities of the choice ${\bf a}$ for the setting of $A$ and ${\bf b}$ for the setting of $B$ would depend on  $[\Psi_{A},\lambda_{A}]$ and  $[\Psi_{B},\lambda_{B}]$ respectively, and they would be uncorrelated. These choices would also  exhibit no correlations with the state vector $\psi$ and the value  $\lambda$ accounting for the state of the composite microsystem. In no sense one can say that the choice of the setting at $A$ is influenced either by the setting at $B$ or by the preparation of the system. It is true that within the considered enlarged scheme the settings are predetermined by the initial conditions, but, due to the absence of correlations, to claim that the choice of the settings influences the whole experiment  would require to make the same claim for all classical physics, something that nobody would take seriously~\footnote{Actually, the argument would boil down to the statement that in a deterministic theory there is no space for free will. To grasp the real meaning of such a position one might consider the claim that, within the Maxwell scenario, since the theory is perfectly deterministic, one cannot assume that an experimenter can freely open or close a switch  of an electric device.  Such a choice is certainly not fully ruled only by electromagnetic laws (in spite of the fact that the nervous transmission involves electromagnetic phenomena) and might be thought of as calling into play also Newtonian mechanics, which, in turn, is fully deterministic. Thus, what the argument amounts to is to accept that the initial conditions, let us say at the big bang, determine in advance (in accordance with the laws of mechanics and electromagnetism) whether the button will be switched on or not. It goes without saying that taking such a position, even though perfectly legitimate, will make the whole scientific enterprise meaningless.}.

\section{5. FREE WILL AND THE NON-SIGNALLING CONDITIONS}

As it emerges from our analysis, the key problem of the present debate is how to impose to a theory the condition that the settings can be chosen freely by the experimenters. We do not intend to tackle here this extremely delicate point in its generality. We think we have already made clear that the way in which Colbeck and Renner have embodied this request in their approach is not justified and, to a large extent, it amounts to assume what they actually want to prove.

From this point of view it is useful to call the attention of the reader on the fact that one might choose other ways, in place of imposing condition (1), to express formally the free will of the experimenters. Just to present an illuminating example we mention that, in a recent paper \cite{ghirardi2}, we have formulated the request of free will by  conditions, denoted by FW, which make reference exclusively to the fact that the two observers can independently choose their settings:
\begin{equation}
P_{A|B\Lambda}=P_{A},\;\;P_{B|A\Lambda}=P_{B}.
\end{equation}
This constraint implies the relevant factorization $P_{AB\Lambda} = P_A P_B P_{\Lambda}$, which means that the two experimenters can freely and independently choose which observable to measure. Moreover, it implies that the hidden variables (or, more in general, the ontic state) cannot influence these choices (compare with the discussion of Section 4). Our request FW is not expected to represent the correct mathematical expression of the free will, but just to provide concrete evidence that assumption FR is stronger than necessary.

Notice that the assumption FW does not involve a third party, and its measurement setting $C$. Actually, the authors of~\cite{colbeck} remark that the additional information supplementing the state vector
\begin{quote}
{\it [...] must be static, that is, its behavior cannot depend on where or when it is observed [...] so, we can consider
the case where its observation is also space-like separated from the measurements specified by $A$ and $B$.}
\end{quote}
Differently from Colbeck and Renner, to make more clear our critical remarks on the assumption FR, we take this statement as an independent assumption, denoted as ST, and mathematically expressed as
\begin{equation}\label{ST}
    P_{CZ|ABXY} = P_{CZ}.
\end{equation}

There is another striking difference between the conditions FR and FW: the former involves the random variables
$X$ and $Y$, the latter does not. We believe that it is important to avoid this dependence, since
the extra variables $X$ and $Y$ could bring spurious correlations, completely independent from the free will assumption.

To better clarify this idea, let us assume that we are interested in the free choice of $A$, and let us raise the following question: if $P_{A|BCYZ} \ne P_A$,
can we conclude that $A$ cannot be freely chosen? We do not think that this is the case. For
instance, we might suppose that the two Stern-Gerlach apparatuses located at the opposite
wings of the experiment could superluminally communicate, at a suitable finite speed. In
the rest frame of the two experimenters, the communication happens after the free choice
of $A$ and $B$, and before the generation of the outputs $X$ and $Y$ . In this way, correlation
between $A$ and $Y$ might be produced. Even though it is highly implausible that a physical
process as the one just mentioned has any physical meaning, its consideration serves the purpose of making clear that
the request $P_{A|BCYZ} \ne P_A$ does not forbid that $A$ and $B$ can be freely and independently chosen. In other words,
as previously remarked, we believe that a good mathematical formulation of the free measurements choice should
not automatically reject situations where free will and superluminal communication coexist.
This is not the case when the FR assumption is considered, since it implies the non-signalling constraints (\ref{NS}),
as stated in Section 1.

In Ref.~\cite{ghirardi2},  we have considered convenient to write non-signalling conditions in a way which is independent from the additional information on the ontic state
and the setting of its measuring device, expressed by:
\begin{equation}\label{NS2}
    P_{X|AB} = P_{X|A}, \qquad P_{Y|AB} = P_{Y|B}.
\end{equation}
Conditions (7) are weaker than relations (\ref{NS}), still they fully express the impossibility
of superluminal communication in the standard EPR scenario. In the following, we refer to Eq. (\ref{NS2})
as the NS assumption.

As proven in~\cite{ghirardi}, the relation between FR, FW, ST and NS is given by
\begin{equation}\label{rel}
    FW \wedge NS \wedge ST \Rightarrow FR,
\end{equation}
and, if the ontic state $\Lambda$ is fully accessible, also the inverse implication holds,
\begin{equation}\label{rel2}
     FR \Rightarrow FW \wedge NS \wedge ST.
\end{equation}
Therefore, in the case considered by Colbeck and Renner, the FR assumption is equivalent to the logical
conjunction of our assumptions FW of free will, staticity ST of the ontic state, and impossibility of superluminal
communication, NS. This result proves our statement that FR is more than the free choice assumption, and
shows that, at least when the ontic state is (partially or totally) unaccessible, negation of FR does
not necessarily imply absence of free will. It might depend on a violation of free will, on the fact that
the additional information on $\Lambda$ is not static, or on a violation of the impossibility of superluminal communication. Of those, the second
condition appears the easiest to digest: why the extra information on $\Lambda$, which is complementing
the information provided by the state vector should be static, when the state vector itself, and the
measurement procedure involved in its preparation, are not space-like separated with respect to $A$
and $B$?

Therefore, the completeness argument presented by Colbeck and Renner has not the claimed generality,
although it is formally correct. In particular, with reference to  Bohmian mechanics, the
statement that measurement settings cannot be freely chosen is not justified. If we
assume that $\Lambda$ is fully accessible (that is, that all positions
are known) it is a well known fact that the theory allows superluminal communication.
On the other hand, the additional information on the ontic state is not static
(in fact, in the case under consideration, it should be a partial information on the positions, which are distributed according to
$\psi$, which in turn is certainly a non-static quantity). In both cases, violation of FR does not
require lack of free will. In Ref.~\cite{ghirardi2} we have been able to work out an interesting completion
of quantum mechanics, precisely in the spirit suggested by Leggett.

\section{6. AN IMPROPERLY FORMULATED NO-GO THEOREM}
To make clearly our argument we mention the statements on which we fully agree with the authors: i) The fact that the locality assumption by Bell (which we will denote as B-Loc in what follows) amounts to the request that $P_{XY|ABZ} = P_{X|AZ} \, P_{Y|BZ}$, provided one assumes that $Z$ represents the maximal specification which is in principle possible of the state of the system; ii) The statement that B-Loc is equivalent to the logical conjunction of the requests of (PI) and (OI); and iii) The proof that their way of imposing FR implies PI. So, the paper by Colbeck and Renner is formally correct and the conclusions correctly follow from the assumptions. But the crucial point is that, as we believe of having made clear, the conditions (1) are, by no means, physically necessary and appropriate.

This leads us quite naturally to establish a parallelism between the celebrated and controversial proof of von Neumann concerning the impossibility of deterministic completions of the theory and the argument of Colbeck and Renner. Von Neumann, by making the (quite natural) assumption that any completion of quantum mechanics had to respect the linear relations between the values of quantum observables has derived, in a correct mathematical way, the proof that no deterministc completion of quantum mechanics is possible.   All of us know very well that it has been just J. Bell who has made perfectly clear that, in spite of its mathematical correctness, von Neumann's proof was not physically significative because it was based on a physically not necessary assumption.

The present situations shares some aspects with the one we have just mentioned. Colbeck and Renner have been guided by the correct idea that the requests of free will has to be imposed to any completion of quantum mechanics. With this in mind, they have translated this request in a precise mathematical condition - which has nothing to do with the one advanced by von Neumann - i.e. their assumption (1). From it they  have correctly derived the conclusion which generalizes the one of von Neumann, i.e. that no completion whatsoever (deterministic or stochastic) of quantum mechanics is possible. One can understand the reasons which have led Colbeck and Renner to resort to condition (1) to express FR, i.e. the fact that it straightforwardly implies the non-signaling conditions which forbid superluminal communication between observers. However, just as in the case of von Neumann, the mathematical constraints they have chosen to impose to the class of theories they have considered are not logically necessary and too strict since one can easily guarantee the impossibility of superluminal communications without resorting to condition (1).

\section{7. CONCLUSIONS}
Our conclusions should be clear to all who have followed the previous argument: the paper by Colbeck and Renner is based on  a too strong and not physically and epistemologically convincing assumption of free will which  trivially implies that all  theories they take into account must satisfy the PI conditions. By analyzing a large class of hidden variable theories, we have made clear that the very starting assumption of  Colbeck and Renner to formalize the free will request is too and unnecessarily strict. Even though we have mainly confined our attention to  deterministic nonlocal hidden variable completions of quantum mechanics we have made clear the reasons for which  the basic claim of Ref.~\cite{colbeck}, that {\it no extension of quantum theory can have improved predictive power}, is not correct when one assumes that at least a part of the parameters completing the theory are not accessible.

\section*{Appendix: An oversimplified completion of quantum mechanics}
We present here an extremely simplified model of a deterministic hidden variable theory, which is a more elegant and symmetric (from the point of view of its nonlocal features) reformulation of Bell's famous example~\cite{bell} of a nonlocal deterministic hidden variable theory for the correlations of the singlet state. Such a model has been recently discussed~\cite{ghirardi2} by the two of us, with reference to a proposal put forward by Leggett~\cite{leggett}. Exactly as Bohmian mechanics, this
model satisfies the aforementioned assumption QM, but it fails to fulfill FR, as we shortly prove.

The model goes as follows. One considers the system of two identical spin $1/2$ particles in the singlet state, which at a given time $t$ (in a given reference frame) are confined to two far away space regions and can be subjected to experiments aimed to ascertain the value of their spin components  along two freely chosen space directions ${\bf a}$ and {\bf b}. The system is characterized by the assignement of the state vector $\psi$ (which in our case is the singlet) and by a hidden variable $\lambda$, which is a unit vector in the three dimensional real space and is assumed to be uniformly distributed on the surface of the unit sphere. One then assumes:
	\begin{itemize}
\item In the case in which only one of the subsystems is subjected to a measurement the value taken by $\sigma^{(1)}\cdot{\bf a}$ is given by $A_{\psi}({\bf a},{\bf b},\lambda) = {\rm sgn}({\bf a}\cdot\lambda)$, and similarly for $\sigma^{(2)}\cdot{\bf b}$, with outcome $B_{\psi}({\bf a},{\bf b},\lambda) = - {\rm sgn}({\bf b}\cdot\lambda)$.
\item On the contrary, if two measurements are performed, the assignment of $\lambda$ uniquely and nonlocally determines the  outcomes as follows: $A_{\psi}({\bf a},{\bf b},\lambda) = {\rm sgn}({\hat{\bf a}}\cdot\lambda)$ and $B_{\psi}({\bf a},{\bf b},\lambda) = - {\rm sgn}({\hat{\bf b}}\cdot\lambda)$, where the vectors $\hat{\bf a}$ and $\hat{\bf b}$ lie in the plane identified by ${\bf a}$ and ${\bf b}$, and are obtained by these vectors by rotating them in such a way that they are still symmetrically disposed with respect to the bisector of the angle $\omega$ (with $0\leq\omega\leq\pi$) between ${\bf a}$ and ${\bf b}$, and form an angle $\hat{\omega}$ satisfying, as in the case of Bell's model, $\hat{\omega} = \pi \sin^{2}\frac{\omega}{2}$. Notice that $\hat{\omega}\leq\omega$ when $\omega \leq \pi/2$, and $\hat{\omega} \geq \omega$ when $\omega> \pi/2$.
\end{itemize}

By using the above relations one can easily prove that the model, when the appropriate average over $\lambda$ is performed, is predictively equivalent to quantum mechanics. Stated differently, it represents, just as Bohmian mechanics, a precise extension (even though limited to the singlet state) of quantum mechanics.

In what follows, we will consider the case in which $X$ is the value of the observable $\sigma^{(1)} \cdot {\bf a}$, $A$ is the space of the settings ${\bf a}$, $B$ is the space of the settings ${\bf b}$ for the observable $\sigma^{(2)}\cdot{\bf b}$, $C$ stays for the space of the hidden variable  and $Z = \lambda$ for a precise value of the hidden variable itself. However, the reader will not encounter any difficulty in understanding the argument keeping in mind these specifications, even though, for simplicity, we will continue to use the generic expressions for the conditional probabilities.

We consider $P_{XZ|AC}$, and we take into account that this probability is different from zero only for $X = \pm 1$ and takes the value $+1$ when $ X = +1$ and ${\bf a} \cdot \lambda > 0$ and the value $0$ when $X = +1$ and ${\bf a} \cdot \lambda < 0$ (the two values must be exchanged when the probabilities refer to the outcome $-1$).
We can now evaluate $P_{XZ|ABC}$. From the prescriptions given above, for some values of ${\bf a}$ and $\lambda$, it is  possible to  choose the setting ${\bf b}$ in such a way to make $P_{XZ|ABC} = 0$ when, for the considered $\lambda$, $P_{XZ|AC} = 1$. Actually, for simplicity let us align $\lambda$ with the $z$-direction and suppose that the  angle between ${\bf a}$ and $\lambda$ is $\theta \leqslant \pi/2$, as shown in  Fig. 1. Let us now consider the direction ${\bf a_{\perp}}$ and let us choose ${\bf b}$ lying between ${\bf a_{\perp}}$ and ${\bf -a}$. Since the angle $\omega$ between ${\bf a}$ and ${\bf b}$ is greater than $\pi/2$, according to the rules governing  our model, ${\hat{\bf a}}$ is rotated, in the clockwise direction, of an amount
\begin{equation}
\frac{\hat{\omega} - \omega}{2} = \frac{\pi}{2} \sin^{2}\frac{\omega}{2} - \frac{\omega}{2}
\end{equation}
with respect to ${\bf a}$. If this amount is greater than $\frac{\pi}{2} - \theta$, i.e., if $\theta$ satisfies the relation
\begin{equation}\label{cond}
\theta > \frac{\pi}{2}\cos^{2}\frac{\omega}{2} + \frac{\omega}{2},
\end{equation}
the quantity $\hat{\bf a} \cdot \lambda$ has opposite sign with respect to ${\bf a} \cdot \lambda$ and, correspondingly, $P_{XZ|ABC} = 0$. It turns out that, when $\pi/2 < \omega < \pi$, condition (\ref{cond}) is always compatible with $\theta \leqslant \pi/2$, and then the model does not respect the non-signaling constraint imposed to any extension of quantum mechanics in Ref.~\cite{colbeck}. Therefore, it does not satisfy Eqs.~(1).
\begin{figure}
\begin{center}
 \includegraphics[width=5cm]{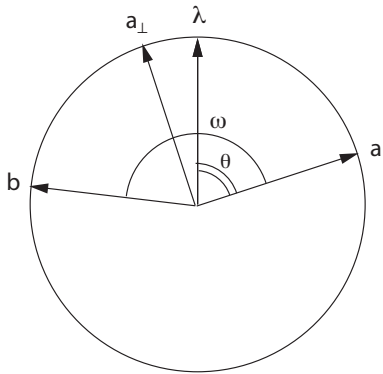} \\
 \caption{\footnotesize Values of ${\bf a},{\bf b}$ and $\lambda$ for which the non-signaling constraints are violated.}
\end{center}
\end{figure}

One might object that our example (which holds only for a specific state $\psi$) is too simplified to allow general conclusions of the sort we have drawn. However, such an objection would be totally out of the target. In fact, anybody familiar with Bohmian mechanics (the paradigmatic example of a hidden variable extension of quantum mechanics) knows very well that, in the case of two far away spin $1/2$ particles in an entangled state, a situation which mirrors perfectly the one we have presented, occurs quite naturally. Typically, for a given state vector $\psi$, for  given values of the positions of the particles (the hidden variables) and for some fixed choice of the setting of the apparatus in $A$, an observer in $B$ can  perform an appropriate choice for the setting of his apparatus in such a way that $P_{XZ|ABC}$ differs from $P_{XZ|AC}$ (obviously here $C$ stays for the positions of the particles and $Z$ for their precise values). So, also in this case, Eqs. (1) are violated.



\begin{thebibliography}{99}

\bibitem{colbeck} Colbeck R. \& Renner R., {\it Nat. Commun.} 2:411 doi: 10.1038/ncomms1416 (2011).

\bibitem{ghirardi2} Ghirardi G.C. \&  Romano R., {\it Onthological models predictively inequivalent to quantum theory}, arXiv:1301.2695, 2013 (accepted by Physical Review Letters)

\bibitem{leggett}  Leggett A.J., {\it Found. Phys.} {\bf 33}, 1469 (2003).

\bibitem{bell}  Bell J.S., {\it Physics} {\bf 1}, 195 (1964).

\bibitem{fine} A. Fine, {\it Phys. Rev. Lett, }{\bf 48}, 291 (1982).

\bibitem{shimony2}  A. Shimony, {\it Controllable and Uncontrollable Non-Locality} in {\it Proc. Int. Symp. Foundations of Quantum Mechanics}, Tokyo, 1983, pp. 225-230.

\bibitem{shimony} A. Shimony, M.A. Horne and J.F: Clauser, Epistemological Letters, F. Bonsack, ed., {\bf 9} 1976, reprinted in {\it Dialectica} {\bf 39}, 97 (1985).

\bibitem{norsen} T. Norsen,  arXiv:0707.0401, 2011.

\bibitem{bell2} J.S. Bell, {\it Journal de Physique}, suppl. au no.3, Tome 42 (1981).

\bibitem{bell3} J.S. Bell, {\it La nouvelle cuisine} in {\it Between Science and Technology}, A. Sarlemijn and P. Kroes eds. Elsevier Publishers, 1990.

\bibitem{ghirardi} G.C. Ghirardi and R. Romano, {\it Comment on ``Is a system's wave function in one-to-one correspondence with its elements of reality?''} arXiv:1302.1635

\end{thebibliography}
\end{document}